# Controlling many-body states by the electric field effect in a two-dimensional material

L. J. Li[1,2,3]*, E. C. T. O'Farrell[1,2]*, K. P. Loh[1,3], G. Eda[1,2,3], B. Özyilmaz[1,2] & A.H. Castro Neto[1,2,#]

[1]Centre for Advanced 2D Materials and Graphene Research Centre, National University of Singapore, 117546, Singapore.

[2]Department of Physics, National University of Singapore, 117542, Singapore.

[3]Department of Chemistry, National University of Singapore, 117543, Singapore.

*These authors contributed equally to this work.

# Corresponding author: phycastr@nus.edu.sg .

To understand the complex physics of a system with strong electron–electron interactions, it is ideal to control and monitor its properties while tuning an external electric field applied to the system. Indeed, complete electric-field control of many-body states in strongly correlated electron systems is fundamental to the next generation of condensed matter research and devices[1–3]. However, the material must be thin enough to avoid shielding of the electric field in the bulk material. Two-dimensional materials do not experience electrical screening, and their charge carrier density can be controlled by gating. 1T-TiSe$_2$ is a prototypical two-dimensional material that shows charge density wave (CDW) and superconductivity in its phase diagram[4], presenting several similarities with other layered systems such as copper oxides[5], iron pnictides[6], and crystals of rare-earth and actinide atoms [7]. By studying 1T-TiSe$_2$ single crystals with thicknesses of 10 nanometres or less, encapsulated in two-dimensional layers of hexagonal boron nitride, we achieve unprecedented control over the CDW transition temperature, tuned from 170 kelvin to 40 kelvin, and over the superconductivity transition temperature, tuned from a quantum critical point at 0 kelvin up to 3 kelvin. Electrically driving TiSe$_2$ over different ordered electronic phases allows us to study the details of the phase transitions between many-body states. Observations of periodic oscillations of magnetoresistance induced by the Little–Parks effect show that the appearance of superconductivity is directly correlated to the spatial texturing of the amplitude and phase of the superconductivity order parameter, **corresponding to a two-dimensional matrix of superconductivity. We infer that this superconductivity matrix is supported by a matrix of incommensurate CDW states embedded in the commensurate CDW states. Our results show that spatially modulated electronic states are fundamental to the appearance of two-dimensional superconductivity.**

The charge carrier density— or equivalently, the Fermi energy—strongly controls phase transitions in correlated systems. Traditionally, charge carrier density can be controlled by doping, that is, by chemical modification of the material. Unfortunately, the alteration of the system's chemical composition leads to the unavoidable introduction of disorder. In strongly correlated systems, owing to their exponential sensitivity to the local electronic environment, disorder can have a profound impact that masks the intrinsic many-body behaviour[8]. Hence, there is a growing need to change the charge carrier density of strongly correlated systems without chemical means. The application of an electric field is one of the cleanest ways to address many-body states since it is intrinsically homogeneous. However, electric fields are screened by the bulk material in three-dimensional metals, making its use difficult.

The Fermi energy not only controls the number of electric carriers but also the screening of external electric fields and internal electron–electron interactions[9]. In two-dimensional (2D) systems the electrons move in a plane while the electric field propagates in three-dimensional space. Hence, 2D electrons are unable to screen electric fields, external or their own. Therefore, we chose to work with a 2D material, $TiSe_2$, of nanometre-scale thickness, and we used an ionic gel electrolyte gate to apply the electric field. In addition, the flake of $TiSe_2$ was encapsulated by a 2D dielectric, hexagonal boron nitride, to avoid external disorder and chemical oxidation and degradation caused by both air and the electrolyte.

Electrical transport measurements under electric-field-induced doping enabled us to construct the phase diagram shown in Fig. 1. Electron doping suppresses the CDW transition from 170 K to 40 K and superconductivity appears with a dome that peaks at 3 K. We show that the emergence of superconductivity is directly associated with the inhomogeneous electronic states that correspond to a periodic structure of the amplitude and phase shifts of the superconductivity order parameter. This periodic structure must be stabilized and pinned to the lattice, so we can infer the presence of an incommensurate CDW (ICDW) matrix surrounding commensurate CDW (CCDW) regions.

TiSe$_2$ nanosheets with thicknesses of 10 nm or less were prepared by mechanical exfoliation of a high-quality single crystal (Extended Data Fig. 1). The device fabrication and measurement details are described in the Methods and the Supplementary Information. In Fig. 2a we sketch the electric-field double layer transistor device used in our experiments, Fig. 2b shows a typical top-gate sweep at 285 K and the variation of the electron density as measured by the Hall effect (see Methods and Extended Data Fig. 2a, b). Using an electrolyte top gate and an electrostatic doped-Si bottom gate we could control the electron density up to about $10^{15}$ cm$^{-2}$ and thereby explore the phase diagram of this 2D material.

Variation of the charge carrier density leads to strong variations of the electrical resistance of the device as shown in Fig. 2c, d. At low charge carrier densities, one can clearly see a peak in the resistivity versus temperature. The CDW transition temperature[10], $T_{CDW}$, corresponds to the inflection point of the resistance and was also measured by using the Hall effect (Extended Data Fig. 2c), to detect the reconstruction of the Fermi surface. On increasing the charge carrier density, $T_{CDW}$ decreases from 170 K to 40 K before becoming undetectable around $n = 7.5 \times 10^{14}$ cm$^{-2}$.

On increasing the electron density we observe the superconductivity state as shown in Fig. 2d. The superconductivity transition temperature, $T_C$, increases from 0 K at the quantum critical point (QCP) at $n = 1.2 \times 10^{14}$ cm$^{-2}$ up to approximately 3 K at an optimal density of $n = 7.5 \times 10^{14}$ cm$^{-2}$. We note that this is exactly the density at which the CDW signal vanishes, indicating a scenario of two competing orders[11]. A further increase in density suppresses $T_C$, giving rise to the formation of a superconductivity dome, as shown in Fig. 1 together with representations of the inferred structure in each region of the phase diagram (discussed below).

When the superconducting coherence length $\xi(T)$ becomes larger than the sample thickness close to $T_C$ we expect the material to behave as a 2D system and the superconducting transition is anticipated to be of the Kosterlitz–Thouless (K–T) type with vortex–antivortex unbinding[12]. One of the trademarks of the K–T transition is the broadening of the resistance with lowering of the temperature, as shown in Fig. 2d. For the K–T transition the resistivity is expected to scale with the coherence length as:

$$\xi(T) \approx a\exp\left(-b/\sqrt{T-T_{\text{K-T}}}\right) \qquad (1)$$

where $T_{\text{K-T}}$ is the K–T transition temperature, and $a$ and $b$ are material parameters. The experimental result reproduces this relation close to $T_{\text{K-T}}$, as shown in Fig. 3a for a charge carrier density of $n = 2.67 \times 10^{14}$ cm$^{-2}$. We also observe current–voltage scaling[13] in the superconductivity phase ($V \propto I^\alpha$) with $\alpha = 5$ for $n = 5.9 \times 10^{14}$ cm$^{-2}$ at the lowest temperature (Extended Data Fig. 3a). By fitting to equation (1), the K–T transition temperatures can be extracted for each doping level. In Fig. 3b we show the behaviour of $T_{\text{K-T}}$ close to the QCP as a function of electron density. Quantum critical scaling[14] predicts $T_{\text{K-T}} \propto (n - n_c)^{z\nu}$, where $z$ is the dynamical exponent and $\nu$ is the correlation length exponent. As shown in Fig. 3b and Extended Data Fig. 3b and c, we find $z\nu \approx 2/3$. The same scaling was observed in other systems[15–17], and indicates that the superconductivity transition is of the classical three-dimensional XY or, equivalently, 2D quantum universality class[16,17]. In the absence of a specific screening or dissipation mechanism[18,19], $z$ is expected to be 1, so that $\nu = 2/3$, which implies that our system is in the clean limit by the Harris criterion[20].

The temperature dependence of the sheet resistance can be written as $R_S = R_{S0} + CT^\alpha$ (where $R_{S0}$ is the residual resistance at 3 K), which decreases monotonically with carrier density as shown in Fig. 3c, in accordance with the above conclusion regarding the Harris criteria. In an ordinary metal (or Fermi liquid) we expect $\alpha = 2$, independent of doping. Nevertheless, in Fig. 3c we find $1 < \alpha < 2$ (Extended Data Fig. 4a and b) over the entire phase diagram. Notice that at around $7.5 \times 10^{14}$ cm$^{-2}$, $\alpha \cong 1.5$ extends down to temperatures close to the superconductivity transition, which would seem to indicate the presence of another QCP, owing to suppression of the CDW, inside the superconductivity dome. In what follows we show that this is not the case.

The presence of the competing orders in this 2D system has striking consequences for the electronic transport. In Fig. 4a we show the magnetoresistance as a function of magnetic field for a density of $n = 5.9 \times 10^{14}$ cm$^{-2}$. The magnetoresistance in the superconductivity phase is positive, as expected, but we clearly observe the presence of plateaus and oscillations in the data. By taking the derivative $dR/dB$, in Fig. 4b we observe that these features are temperature-independent and have well defined periods. The periodicity in magnetic field reflects a spatial periodicity given by the cyclotron equation, $\ell_c = (\Phi_0/\delta B)^{1/2}$, where $\Phi_0 = h/(2e) \approx 2{,}068$ T nm$^2$ is the flux quantum and $\delta B$ is the magnetic field periodicity. We have analysed the magnetoresistance data as a function of electron density (or gate voltage). (Magnetoresistance data for other electron densities is shown in Extended Data Fig. 5.) One can see a clear trend in the data: the length scale decreases monotonically with electron density from $\ell(n) \approx 450$ nm at $n \approx 1.3 \times 10^{14}$ cm$^{-2}$ to $\ell(n) \approx 170$ nm at $n \approx 5.9 \times 10^{14}$ cm$^{-2}$.

It is clear that the well defined structure of the magnetoresistance reflects spatial fluctuations of the electronic pairing. A consistent explanation for these features is based on the Little–Parks effect[21], whereby Cooper pairs are constrained to move in loops in the material—that is, pairing is local and constrained to well defined regions. The length scale we observe is associated with the trapping of magnetic flux quanta by the Cooper pairs. The existence of such a superconductivity matrix over a range of temperature and charge carrier density in a single crystal is remarkable, leading us to suggest that the superconductivity matrix must be pinned and stabilized by an underlying matrix of inhomogeneous electronic states. Fluctuations of an underlying charge or spin order parameter have led to the discovery of superconductivity in a wide range of systems; the suppression of the CDW transition from 170 K to 40 K, concomitant with the appearance of superconductivity and non-Fermi-liquid behaviour, strongly indicates that the CDW plays a part.

We now consider how local variations of the CDW can stabilize the superconductivity matrix. The CDW corresponds to a spatial modulation of the charge density $\delta\rho = \Delta(\boldsymbol{r})e^{-i[Q(r)\cdot r]}$, where $\Delta$ is the CDW order parameter, r is the position in the 2D plane and Q is the CDW ordering vector. On the basis of symmetry alone, the Ginzburg–Landau free energy for $\Delta$ can be written as[22]:

$$F = \int d\boldsymbol{r} \left\{ a(\Delta)^2 + b(\Delta)^3 + c(\Delta)^4 + \frac{1}{2m^*Q^2}\left[ |Q(\nabla - iQ)\Delta|^2 + \kappa |Q \times \nabla \Delta|^2 \right] \right\} \quad (2)$$

where $a$, $b$, $c$, $m^*$ and $\kappa$ are phenomenological parameters that determine the energy scale for the spatial variation of $\rho$. Variations in the amplitude $\Delta$ are energetically very costly since the CDW has to be locally destroyed. However, variations in the CDW phase are energetically allowed and can be expressed as:

$$\Delta(\boldsymbol{r}) = \Delta_0 \exp\left\{ i\left[ \frac{\mathbf{K}}{2}\mathbf{r} - \theta(\mathbf{r}) \right] \right\} \quad (3)$$

where $\Delta_0$ is the CDW order parameter in the uniform CCDW, $K$ is a reciprocal lattice vector in the direction of $Q$, $\theta(r)$ is a spatially varying phase and we have included the known CCDW wavevector for TiSe$_2$ (1/2, 1/2, 1/2). When $\theta = \pi n$ (where $n$ is an integer) we again have a CCDW, whereas if $\theta(r) = Q.r$ we have an ICDW. For illustrative purposes, we assume that the variation in $\theta(x)$ is one-dimensional in nature and substitute equation (3) into equation (2) to obtain:

$$\delta F = \int dx \frac{1}{2} \left\{ \left[ \partial_x \theta(x) - 1 \right]^2 - g \left[ 1 - \cos 2\theta(x) \right] \right\} \quad (4)$$

where $g$ depends on the Ginzburg–Landau parameters and $x = |K/2 - Q|$ r is the dimensionless length scale. The last equation reflects how the free energy changes locally with the phase. Minimizing equation (4) with respect to $\theta$ we obtain:

$$\frac{d^2\theta}{dx^2} = -2g\sin(2\theta) \quad (5)$$

the differential equation for the pendulum ($x$ plays the part of 'time'), which has periodic solutions with period $\ell_{ICDW} \approx \pi/(g^{1/2}|K/2 - Q|)$. These spatially periodic variations of the phase represent Neél-like domain walls of ICDW between CCDW regions with different phase where $\ell_{ICDW}$ is the thickness of the domain wall. McMillan described[23] how these defects allow a uniform ICDW with slowly varying phase to break apart into domains of CCDW separated by ICDW domain walls that have more rapidly varying phases. The domain wall density is $\pi/Q$ to match the homogeneous ICDW state.

A full solution to this problem in 2D is lacking, but we speculate that domain walls form a periodic matrix illustrated schematically in Fig. 1; blue CCDW regions with constant phases are embedded in a periodic ICDW matrix. We note that a similar structure was observed by scanning transmission microscope measurements of the closely related 1T-TaS$_2$, in which the ICDW state exists at ambient conditions[24]. The self-organizing principle is that repulsive interactions occur between domain walls owing to higher-order terms in the free energy[23]. Therefore the ICDW domains will form a matrix, breaking the CCDW into domains with fixed area, as required by the Little–Parks effect.

As shown by ref. 23, ICDW dynamic phase fluctuations—that is, phonon modes of the ICDW (not the lattice)—can exist in these domain walls. It is conceivable that these ICDW phonons induce superconductivity pairing and localize Cooper pairs in one-dimensional regions of the 2D system. Another intriguing aspect of our results is displayed in Fig. 4d, which shows the point-contact conductance spectra measured at each density, in which we observe a clear zero bias conductance peak (ZBCP) in the superconductivity state. Extended Data Fig. 6 shows its temperature and magnetic field dependence at a density of $n = 2.1 \times 10^{14}$ cm$^{-2}$. ZBCPs are observed in a wide range of unconventional superconductors and are understood to arise by Andreev reflection from a Cooper pairing potential having an internal phase shift of the superconductivity order parameter[25]. These results are therefore in stark contrast to the experimentally determined single-gap $s$-wave superconductivity observed in the Cu-intercalated Cu$_x$TiSe$_2$ (ref. 26). It is unlikely that our 2D samples would develop a superconductivity order parameter that is qualitatively distinct from that of Cu$_x$TiSe$_2$ (for example, $d$-wave). The existence of the ZBCP therefore suggests that, together with the spatial modulation of the superconductivity amplitude (which is demonstrated by the Little–Parks effect), 、there may also be a modulation of the superconductivity phase, although the correspondence between the amplitude variation and the phase variation cannot be determined from our measurements.

The observed state in 1T-TiSe$_2$ bears some similarity to the pair density wave (PDW) superconducting CDW phases. However, further experiments are required to substantiate the PDW hypothesis. Although one-dimensional PDW states have attracted more attention within the context of the cuprate superconductors[27], more general PDW states having phase and amplitude variations in 2D are expected to be possible[28].

The coexistence of CCDW and ICDW was first observed by recent X-ray measurements of TiSe$_2$ at pressures close to where the superconductivity phase was expected[29]. ICDW domain walls with a periodicity along the $c$-axis of ~300 nm were observed, similar to the length scale determined in this experiment. While the periodicity was most pronounced along the $c$-axis, a weak in-plane signal of incommensurability was observed that might correspond to the electronic microstructure observed here in the superconducting order (Abbamonte, P., personal communication, 15 January 2015).

In summary, we studied samples of TiSe$_2$ a few nanometres in thickness and tuned the material through the CDW and superconductivity phases using the electric field effect. This technique allowed us to study in great detail the QCP in the material and classify its universality class. We also identified the interplay between superconductivity and CDW through the formation of an inhomogeneous many-body state which we identify with the localization of Cooper pairs along a matrix of incommensurate dislocations surrounding regions of CCDW. We conjecture that the superconductivity has in its origin in the coupling of the McMillan phonon modes of the ICDW with the electrons. These results open a new era of electric-field tuning of many-body states in condensed matter research.

**Acknowledgements:**

We thank L. Q. Chu, T. H. Ren, J. Y. Tan, J. Wu and H. Schmidt for experimental assistance, and S. Natarajan for assistance in preparing the manuscript. A.H.C.N. acknowledges many discussions with A. K. Geim and a private communication with P. Abbamonte. K.P.L. acknowledges a MOE Tier 1 grant, "2-D crystals as a platform for optoelectronics (R-143-000-556-112)". G.E. acknowledges a National Research Foundation (NRF) Research Fellowship (NRF-NRFF2011-02). B.Ö. acknowledges support by the NRF, Prime Minister's Office, Singapore, under its Competitive Research Programme (CRP award number NRF-CRP9-2011-3), and the SMF-NUS Research Horizons Award 2009-Phase II. A.H.C.N acknowledges the CRP award, "Novel 2D materials with tailored properties: beyond graphene" (NRF-CRP6-2010-05). . All authors acknowledge the NRF, Prime Minister Office, Singapore, under its Medium-Sized Centre Programme.

**Author Contributions:**

L.J.L. performed the growth, characterization of the single crystal, L.J.L. and E.C.T.O'F. performed device fabrication and carried out the measurement, L.J.L., E.C.T.O'F., K.P.L, B.Ö. , and A.H.C.N. analysed the data and wrote the manuscript. All authors commented on the manuscript.

Author Information Reprints and permissions information is available at www.nature.com/reprints. The authors declare no competing financial interests. Readers are welcome to comment on the online version of the paper. Correspondence and requests for materials should be addressed to A.H.C.N. (phycastr@nus.edu.sg).

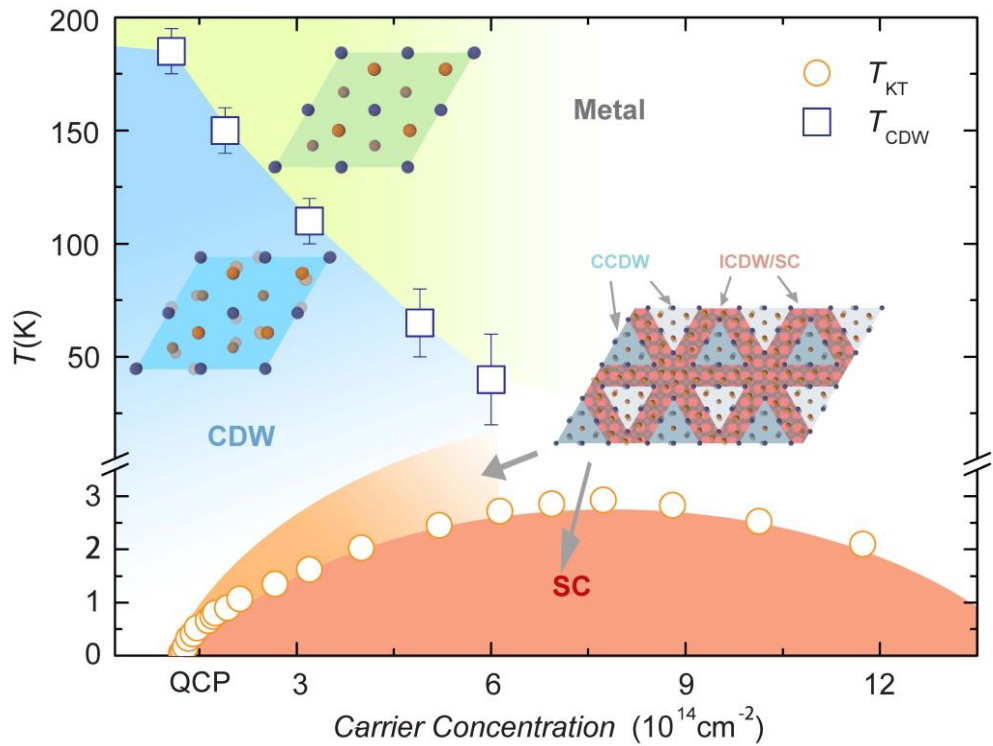

Figure 1 Phase diagram of TiSe$_2$ under electron doping. Circles show $T_{K-T}$ and squares show $T_{CDW}$. The insets show the lattice structure in each phase. In the CDW phases we illustrate the atomic displacements within an enlarged unit cell; in the phase in which ICDW and CDW coexist we schematically illustrate the ICDW domain walls between the CCDW regions as the red region (which we have exaggerated to occupy only a single unit cell). The error bars define the difference in $T_{CDW}$ between the values derived from resistivity versus temperature curves and those derived from carrier density versus temperature curves.

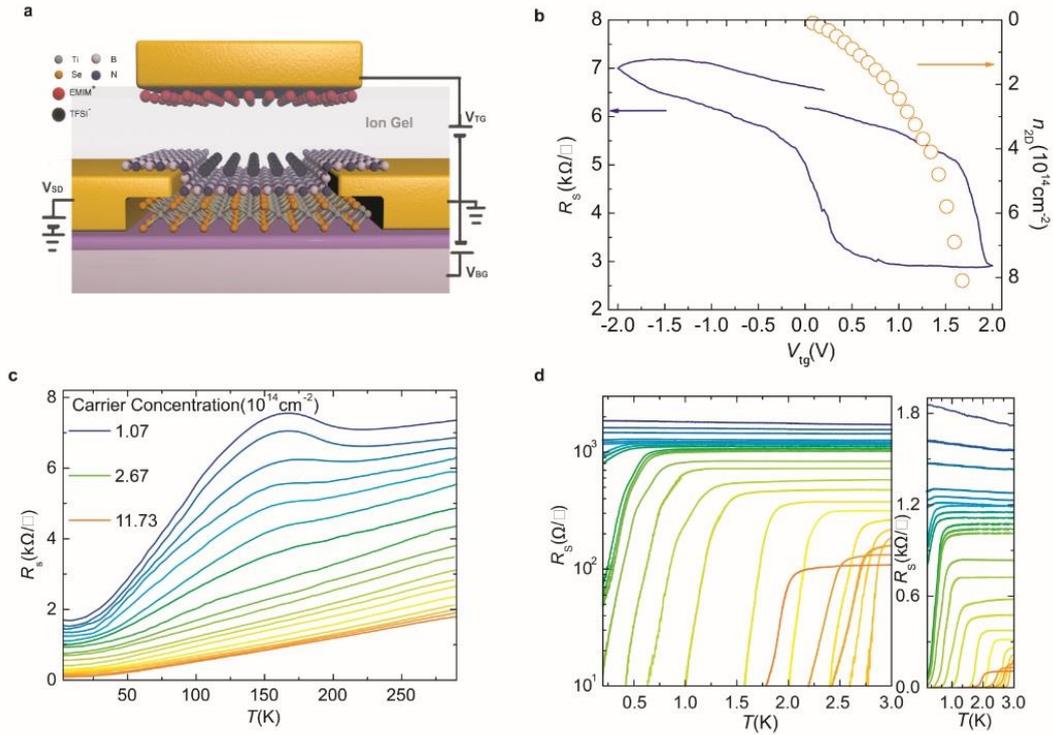

Figure 2 Characterization of the field effect device and the resistance at different doping levels by gating. a, Sketch of the electric-field double layer transistor device. $V_{SD}$, source–drain voltage; $V_{TG}$, top gate voltage; $V_{BG}$, bottom gate voltage. b, Typical device characteristic under ion gel top-gate operation at 285 K. Electrical resistance $R_S$ is shown on the left and electron density $\delta n_{2D}$, measured by the Hall effect at 285 K, on the right. c and d show that at high and low temperatures, respectively, the CDW and superconductivity transitions can be clearly identified.

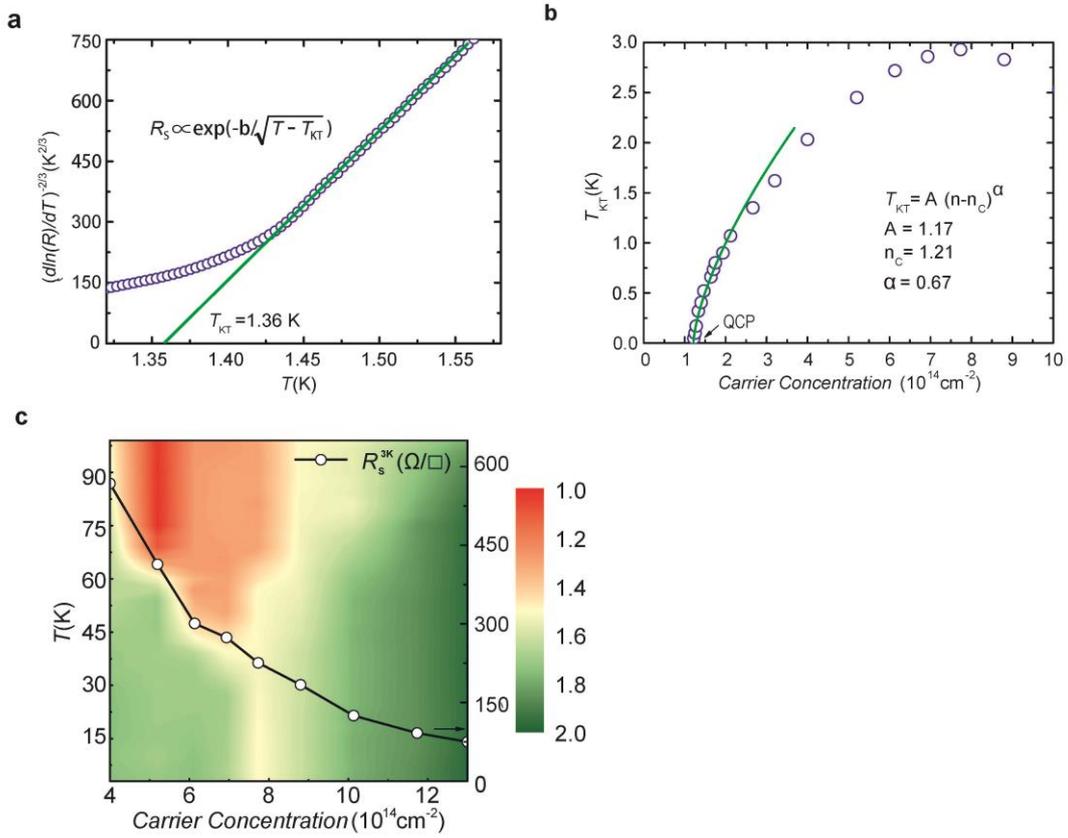

Figure 3 Temperature dependence of the electrical resistance close to the K–T transition. a, Fitting of the resistance to the K–T formula for a carrier density of $2.67 \times 10^{14}$ cm$^{-2}$. b, Behaviour of $T_{\text{K–T}}$ at the QCP with critical exponent $z\nu \approx 2/3$. c, $\alpha$ values derived by fitting the $R$ versus $T$ data from 3 K to 100 K with the formula $R_S = R_{S0} + C \cdot T^{\alpha}$, where $\alpha$ is plotted in colour against both carrier density and temperature. The residual resistance at 3 K versus carrier density is plotted on the right axis.

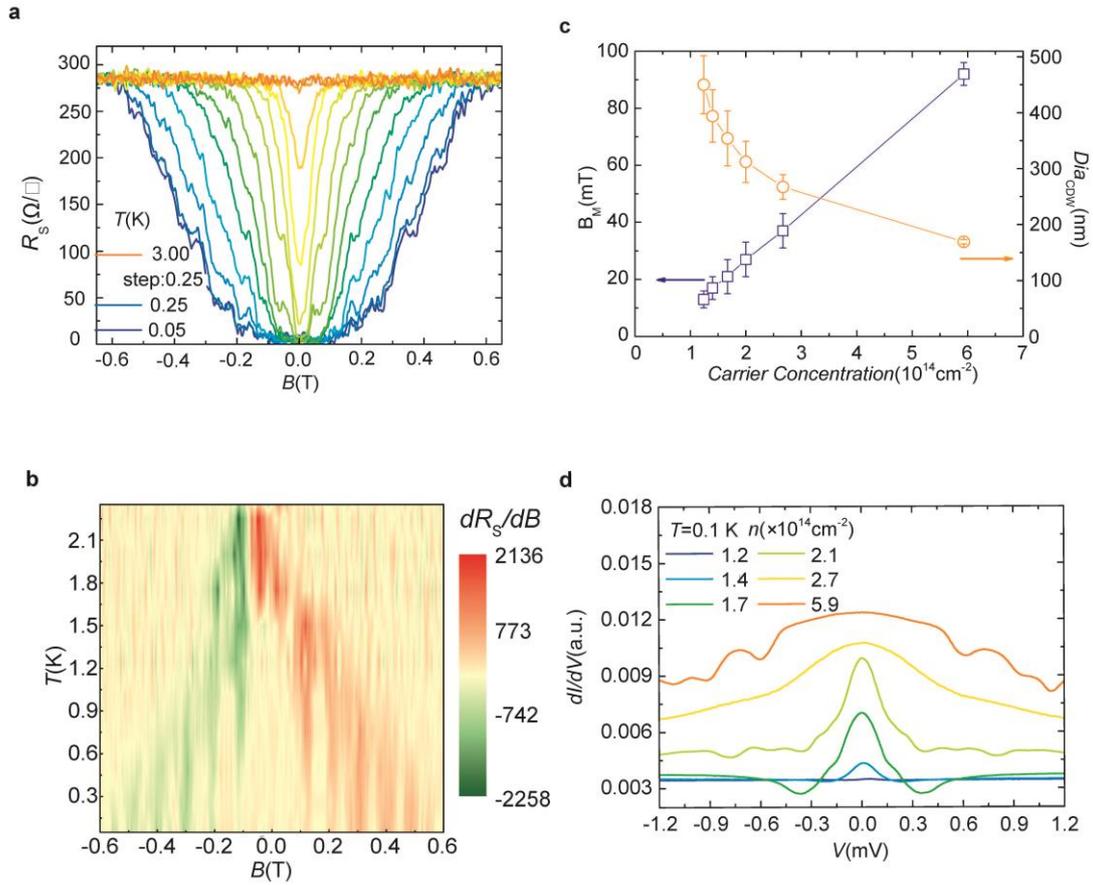

Figure 4 The magnetoresistance for a carrier density of $5.9 \times 10^{14}$ cm$^{-2}$ shows periodic oscillation. a, The magnetoresistance at different temperatures (each step is 0.25 K) in the superconductivity. b, Colour contour plot of the derivative of magnetoresistance d$R_s$/d$B$ versus magnetic field $B$ and temperature $T$, which displays temperature-independent periodic oscillations. c, Derived magnetoresistance oscillating period $B_M$ and the corresponding length scale Dia$_{CDW}$. Error bars define the 90% confidence interval. d, Carrier-density-dependent two-terminal conductance d$I$/d$V$ shows the ZBCP, indicating non-$s$-wave superconducting pair symmetry. a.u., arbitrary units.

**METHODS**

**Crystal growth and quality verification**

TiSe$_2$ single crystals were grown in two steps by the chemical vapour transport method[30]. First, polycrystalline TiSe$_2$ was prepared by mixing high-purity titanium powder (from Alfa Aesar, 99.99%) and selenium powder(from Alfa Aesa, 99.999%) in a stoichiometric ratio and heating the mixture at 800 °C for 3 days in a vacuum-sealed (<10$^{-6}$ torr) silica tube. Second, the polycrystalline powder was loaded into a two-zone tube furnace together with the transport agent $I_2$ at a concentration of 5 mg cm$^{-3}$. The polycrystalline powder was then heated to 670 °C and single crystals of TiSe$_2$ were collected at 600 °C over a period of 10 days.

The quality of the bulk single crystals was confirmed by X-ray diffraction and temperature-dependent Raman spectroscopy[31], as shown in Extended Data Fig. 1a and b. Further energy dispersive X-ray spectroscopy verified the stoichiometric composition of the crystals.

**Device fabrication and characterization**

TiSe$_2$ was exfoliated in a pure argon atmosphere by Scotch tape[32] onto a SiO$_2$ (300 nm)/Si wafer and examined under high-resolution optical microscope. The non-uniformity in thickness can be discriminated by cross-correlation of the colour with atomic force microscope measurements of the height. Flakes with uniform thickness of around 10 nm or less and a long bar shape were selected for the device fabrication. Electrodes for transport measurements were fabricated by standard electron beam lithography techniques using a PMMA positive resist, followed by deposition of Ti (10 nm)/Au (65 nm). Thin crystals (one to three layers) of commercial hexagonal boron nitride were transferred onto the nanosheets within the argon atmosphere[33,34]; the role of hexagonal boron nitride is to protect the TiSe$_2$ from degradation by both oxidation and damage by the electrolyte gate.

Atomic force microscope results show that the surface is clean (as shown in Extended Data Fig. 2b), with a roughness within ±1 nm, which may result from the non-uniform thickness of the TiSe$_2$ flake we mean the submicron areas in AFM images.

Electrical transport measurements were performed in both a $^4$He cryostat and in a $^3$He/$^4$He dilution cryostat. Electrical transport measurements were performed using standard alternating-current (a.c.) lock-in amplifier and direct-current (d.c.) techniques, and resistance against temperature and field measurements were performed using currents of 10–100 nA to avoid Joule heating.

The ion gel solution was prepared by mixing the triblock copolymer poly(styrene-block-methylmethacrylateblock-styrene) (PS-PMMA-PS) and the ionic liquid 1-ethyl-3-methylimidazolium bis(trifluoromethylsulfonyl)imide (EMIM-TFSI) into an ethyl propionate solvent (the weight ratio of polymer to ionic liquid to solvent is 0.7:9.3:90)[35]. After covering the device with ion gel droplets by drop casting, the device as shown in Extended Data Fig. 2a was loaded into the cryostat and kept at room temperature and high vacuum for one hour to remove residual water from the electrolyte. Afterwards, resistance was measured against gate voltage to characterize the capability of the ion gel; a typical electrolyte gate sweep is shown in Fig. 1b.

The carrier density doping by the ionic gate can be derived from the Hall-effect measurement both at high (285 K) and low (3 K) temperature; the former is shown in Extended Data Fig. 2c, and the latter is used to construct the phase diagram since the latter has a better direct correlation with the superconducting dome. Although the hexagonal boron nitride passivation prevents the accumulation of ions directly at the surface of TiSe$_2$, this was not found to reduce the capacity of the gate much, as demonstrated by our results and those of a recent work[36].

**2D superconducting properties and the K–T transition**

As discussed in the main text, the superconducting transition under different fixed perpendicular magnetic fields was measured. Extended Data Fig. 3b shows the magnetoresistance plot for a charge carrier density of $n = 2.67 \times 10^{14}$ cm$^{-2}$. The upper critical field $H_{C2}(T)$ values can be determined from each curve at the intercept of extrapolations from the normal state and the superconductivity state. $H_{C2}(0)$ can be derived by interpolating the plot curve to zero temperature, which gives a value of 450 mT. The superconducting coherence length therefore can be derived from $\xi(0) = \sqrt{H_{C2}(0)/\phi_0}$. The minimum $\xi$ that corresponds to the maximum $H_{C2}$ point is about 22 nm, which is more than twice as large as the thickness of the measured device, indicating that superconductivity in our device is expected to have a 2D character.

For the K–T transition, the current–voltage response becomes nonlinear; a $V \propto I^\alpha$ relation is expected as a result of the vortex–antivortex pair unbinding[37]. Consistent behaviour is also observed in our sample, as shown in Extended Data Fig. 3a.

To confirm the $\nu z$ value that determines the nature of the quantum critical behaviour in our system, we also measured the temperature-dependent superconducting transition under perpendicular magnetic field at a fixed charge carrier density of $n = 2.67 \times 10^{14}$ cm$^{-2}$ (see Extended Data Fig. 3b). By using finite size scaling with the formula $R_S/R_C = F\{(B - B_C) \times T^{-1/\nu z}\}$, where $R_C$, $B_C$ are two fitting parameters, $F$ is an arbitrary function with $F(0) = 1$ (ref. 38), the data are expected to collapse into two sets of lines, with a certain $\nu z$ value. As displayed in Extended Data Fig. 3c, the data collapse for $\nu z \approx 2/3$, which confirms the previous result.

**Magnetoresistance oscillations at other doping levels**

The magnetoresistance oscillation is observed when we sweep a perpendicular magnetic field at different temperature in the superconducting state. From the QCP point $n = 1.2 \times 10^{14}$ cm$^{-2}$ to the near-optimum doping $n = 5.9 \times 10^{14}$ cm$^{-2}$, the oscillations can be observed for all doping levels. However, these oscillations can only be clearly observed for certain temperatures $T_0$ and magnetic fields $B_0$, whereas $T_0$ and $B_0$ values increase with increasing doping. For instance, for $n = 1.3 \times 10^{14}$ cm$^{-2}$, $T_0$ is 0.3 K and $B_0$ is 0.06 T; for $n = 2.7 \times 10^{14}$ cm$^{-2}$, $T_0$ increases to 0.4 K and $B_0$ increases to 0.13 T, as one can see from Extended Data Fig. 4. Although $T_0$ and $B_0$ values as well as the periods of oscillation $\delta B$ increase with doping level, the amplitude of the magnetoresistance oscillation does not monotonically depend on doping levels. We find that the oscillating amplitude for doping levels of $1.3 \times 10^{14}$ cm$^{-2}$ and $5.9 \times 10^{14}$ cm$^{-2}$ is larger than that for other doping levels we measured. One can clearly see more contrast or sharpness for the periodic straight lines in Fig. 4b and Extended Data Fig. 4b than in Extended Data Fig. 4d. The stronger magnetoresistance oscillations at these doping levels could be related to the enhanced Cooper-pair phonon interaction, aroused by strong quantum fluctuation.

**Temperature dependence of the sheet resistance**

We plot the temperature dependence of the sheet resistance between 3 K and 100 K with the doping level ranging from $4 \times 10^{14}$ cm$^{-2}$ to $13 \times 10^{14}$ cm$^{-2}$ as shown in Extended Data Fig. 5a and b. By taking the temperature derivative $d(\log(R - R_0))/d(\log(T))$, $\alpha$ is extracted at each doping as a function of the temperature.

At doping levels away from optimal doping, $7.5 \times 10^{14}$ cm$^{-2}$, we observe Fermi-liquid behaviour at low temperatures below $T_{CDW}$. At the optimal doping level an exponent of 3/2 is observed over a wide range of temperature; this exponent is similar to the one observed in MnSi[39]. As described in the main text, microscopic fluctuations of the order parameters from those of a CCDW to those of an ICDW gives rise to this temperature dependence.

**Point-contact conductance spectroscopy**

Point-contact conductance spectroscopy of the normal-superconducting junction between Au/Ti and TiSe$_2$ was performed by the two-terminal a.c. + d.c. method, whereby the d.c. voltage is modulated with an additional a.c. voltage, such that the derivative d$I$/d$V$ can be measured at the first harmonic by a current preamplifier and standard lock-in amplifier techniques.

The contacts were patterned by standard electron beam lithography using a PMMA positive resist. The development of the resist is performed in air to allow the oxidization of the contact region such that the contacts (despite not being nanoscale) are in the so-called 'soft' contact regime that has been successfully applied to pnictide and cuprate superconductors[40]. In this regime spectroscopic information can be obtained because the transport is primarily through multiple point-like pinholes whose individual dimension is smaller than the mean free path in the contact.

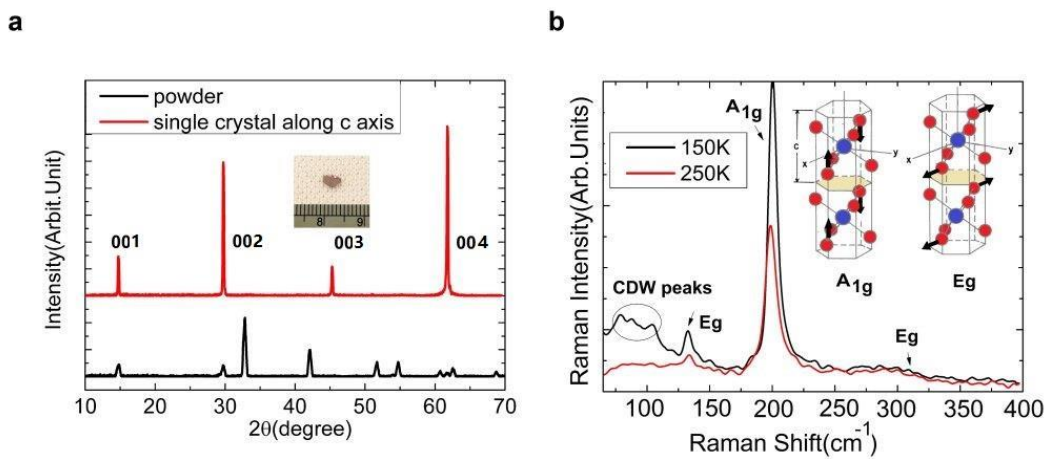

Extended Data Fig. 1 Characterization of the high quality of single-crystal $TiSe_2$. a, X-ray diffraction of both single-crystal and powder $TiSe_2$ sample. The inset shows the as-grown $TiSe_2$ single crystal. b, Raman spectroscopy pattern at both high temperature and low temperature. The two main phonon modes, $E_g$ and $A_{1g}$, are distinct, whereas only below $T_{CDW}$ are the peaks corresponding to CDW phonon mode detectable. The inset displays the unit cell of the $TiSe_2$ lattice and the main phonon mode vectors.

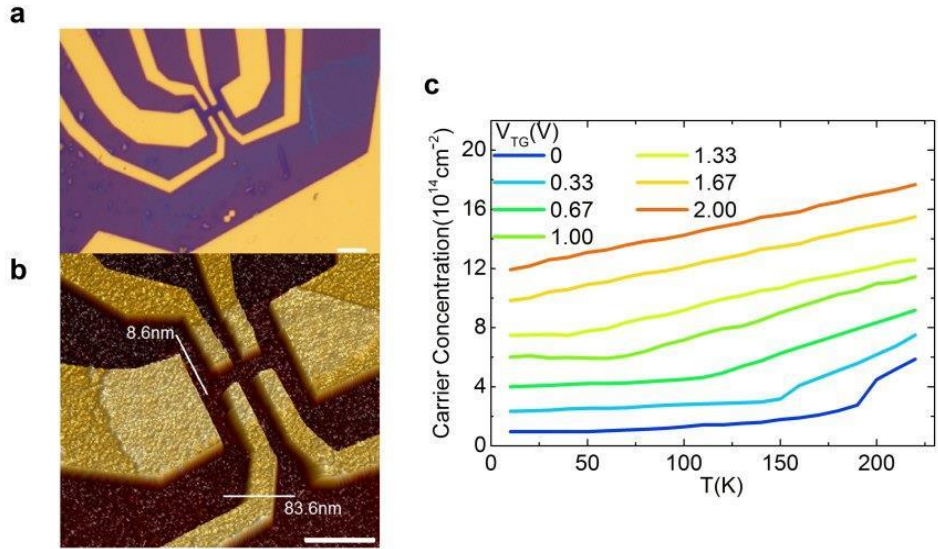

Extended Data Fig. 2 The Hall bar device and its characterization by Hall effect measurement. a, Optical microscope picture. b, Atomic force microscope picture of the Hall bar device. c, Temperature dependence of the carrier density measured by the Hall effect at different top gate voltages, $V_{TG}$. Scale bar, 5 μm.

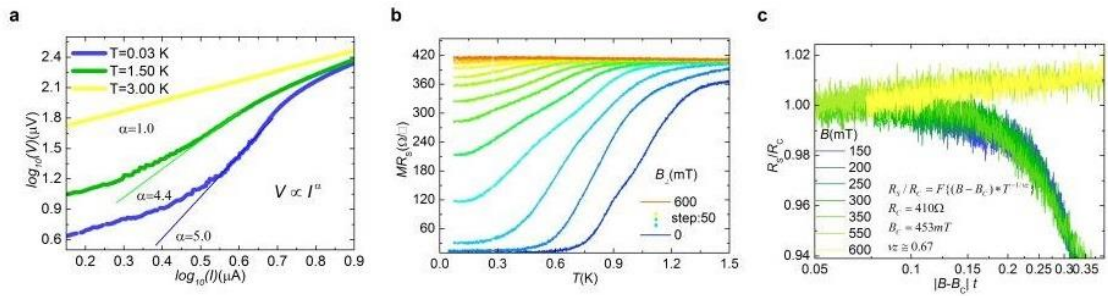

Extended Data Fig. 3 Characterization of the KT transition. a, The current-voltage power-law fit for $n = 5.9 \times 10^{14}$ cm$^{-2}$ at different temperatures is consistent with the behaviour of the 2D K–T transition. b, Temperature-dependent magnetoresistance of the superconducting transition at different fixed perpendicular magnetic fields for $n = 2.67 \times 10^{14}$ cm$^{-2}$. c, The magnetoresistance data in b collapses into two sets of lines by so-called finite size scaling.

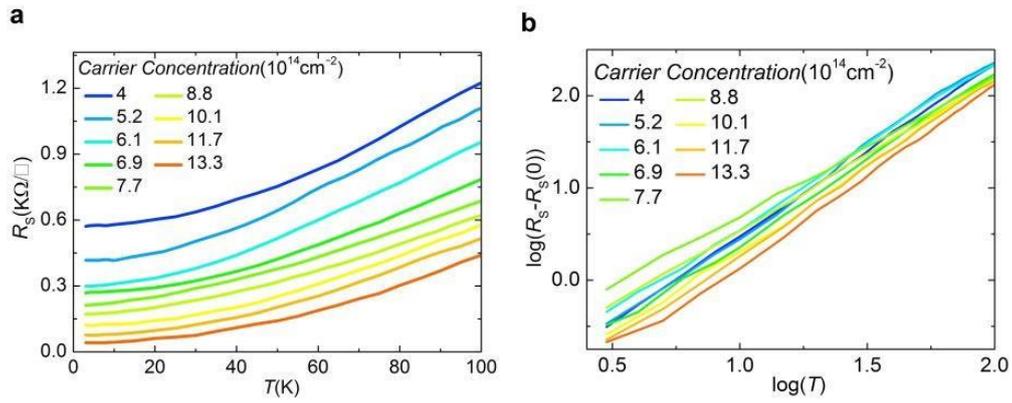

Extended Data Fig. 4 The *R* versus *T* power-law fit indicates the existence of strong quantum fluctuation. a, Temperature dependence of the sheet resistance for different doping levels. b, The data shown in a is plotted in log-log scale.

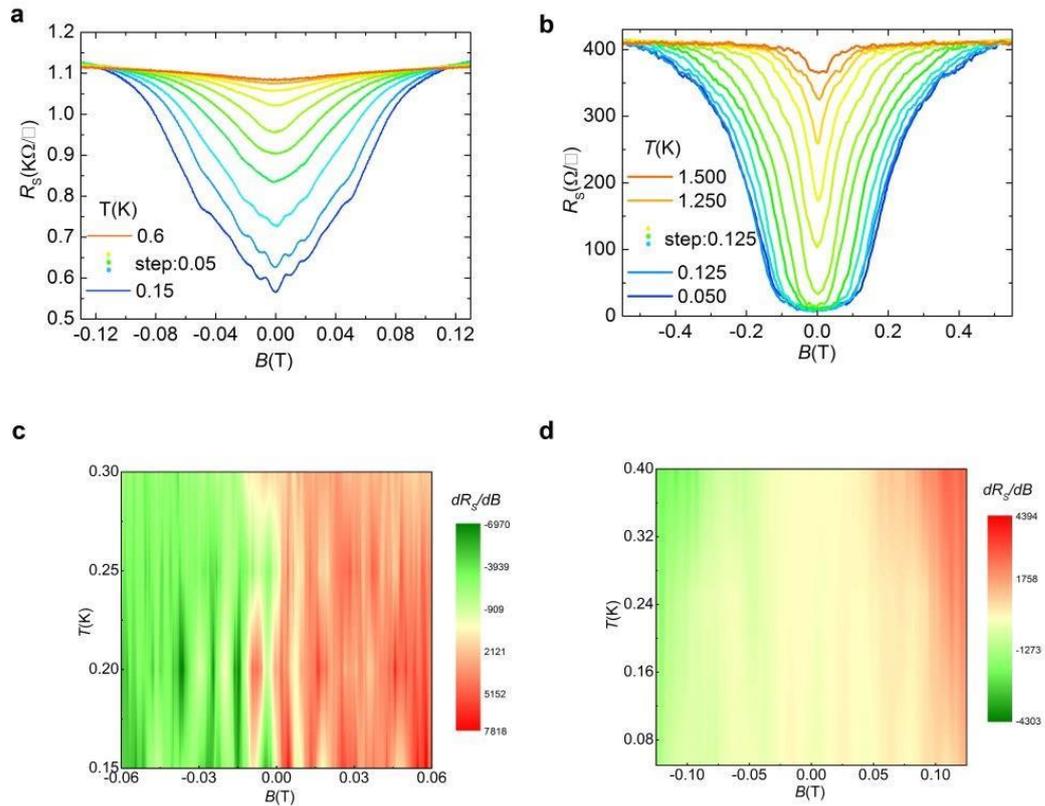

Extended Data Fig. 5 The magnetoresistance oscillation for carrier densitys of $1.3 \times 10^{14}$ and $2.7 \times 10^{14}$ cm$^{-2}$. a, c, Perpendicular magnetic-field-dependent magnetoresistance measured at different temperatures. b, d, Plot of d$R_S$/d$B$ against $B$ and $T$ for $1.3 \times 10^{14}$ cm$^{-2}$ and $2.7 \times 10^{14}$ cm$^{-2}$, respectively.

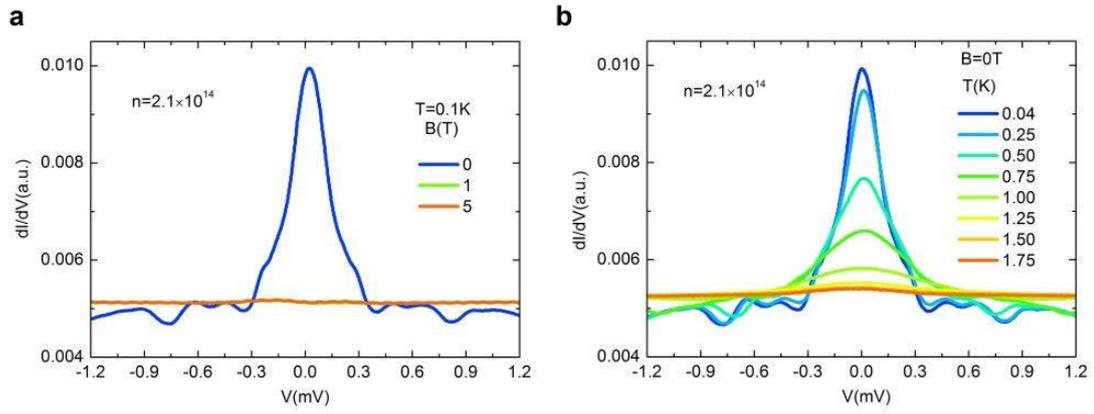

Extended Data Fig. 6 The conductance measured for a carrier density of $2.1 \times 10^{14}$ cm$^{-2}$. a, Magnetic field dependence at 0.1 K. b, Temperature dependence at zero magnetic field.